# Fingerprints of carbon defects in vibrational spectra of gallium nitride (GaN) considering the isotope effect

I. Gamov[1,*], J. L. Lyons[2], G. Gärtner[3], K. Irmscher[1], E. Richter[4], M. Weyers[4], M.R. Wagner[5], and M. Bickermann[1]

[1] *Leibniz-Institut für Kristallzüchtung (IKZ), Berlin, Germany*
[2] *Center for Computational Materials Science, United States Naval Research Laboratory, Washington, DC, USA*
[3] *Institute of Experimental Physics, TU Bergakademie Freiberg, Freiberg, Germany*
[4] *Ferdinand-Braun-Institut (FBH), Berlin, Germany*
[5] *Institute of Solid State Physics, Technische Universität Berlin, Berlin, Germany*
[*]*corresponding author, current affiliation: Technische Physik, University of Würzburg, Würzburg, Germany, e-mail:* ivan.gamov@uni-wuerzburg.de

**Abstract**

This work examines the carbon defects associated with recently reported and novel peaks of infrared (IR) absorption and Raman scattering appearing in GaN crystals at carbon ($^{12}$C) doping in the range of concentrations from $3.2 \times 10^{17}$ to $3.5 \times 10^{19}$ cm$^{-3}$. 14 unique vibrational modes of defects are observed in GaN samples grown by hydride vapor phase epitaxy (HVPE) and then compared with defect properties predicted from first-principles calculations. The vibrational frequency shift in two $^{13}$C enriched samples related to the effect of the isotope mass indicates six distinct configurations of the carbon-containing point defects. The effect of the isotope replacement is well reproduced by the density functional theory (DFT) calculations. Specific attention is paid to the most pronounced defects, namely tri-carbon complexes ($C_N=C=C_N$) and carbon substituting for nitrogen ($C_N$). The position of the transition level (+/0) in the bandgap found for $C_N=C=C_N$ defects by DFT at 1.1 eV above the valence band maximum, suggest that $(C_N=C=C_N)^+$ provides compensation of $C_N^-$. $C_N=C=C_N$ defects are observed to be prominent, yet have high formation energies in DFT calculations. Regarding $C_N$ defects, it is shown that the host Ga and N atoms are involved in the defect's delocalized vibrations and significantly affect the isotopic frequency shift. Much more faint vibrational modes are found from di-atomic carbon-carbon and carbon-hydrogen (C-H) complexes. Also, we note changes of vibrational mode intensities of $C_N$, $C_N=C=C_N$, C-H, and $C_N$-$C_i$ defects in the IR absorption spectra upon irradiation in the defect-related UV/visible absorption range. Finally, it is demonstrated that the resonant enhancement of the Raman process in the range of defect absorption above 2.5 eV enables the detection of defects at carbon doping concentrations as low as $3.2 \times 10^{17}$ cm$^{-3}$.

## 1. Introduction

Carbon is a common contaminant in the epitaxy of GaN crystals, as well as an intentional dopant in GaN-based high-power device technologies. It is frequently used to achieve semi-insulating substrates and buffer layers.[1–6] In *n*-type GaN, carbon doping leads to the formation of compensating acceptors, in which a single carbon atom substitutes for a nitrogen host atom ($C_N$).[7–10] This defect is usually considered as the prevailing carbon defect in *n*-type GaN due to its low formation energy.[11–17] $C_N$ possesses a deep acceptor level (0/-) at ~1 eV above the valence band maximum (VBM) that is in good agreement with the photoluminescence, high temperature conductivity, bulk photovoltaic effect, and charge transfer processes observed in the material.[12,18–20] Deep-level transient spectroscopy (DLTS) studies have supported the finding that there is a (+/0) level of $C_N$ that can act as a hole trap, compensating Mg doping in *p*-type GaN.[21] Two intense IR absorption bands in GaN appearing at 1678 and 1718 cm$^{-1}$ in GaN:C samples were interpreted as antisymmetric vibrational modes ($v_3$) of tri-carbon defects,[22,23] and correlate with a characteristic red photoluminescence band.[24] However, these defects are distinct from the carbon complexes in theoretical reports [10,14,16] and hence the role of the observed tri-carbon defects in the charge balance is not yet understood.

Often carbon is described as an "amphoteric" impurity because its propensity to act as an acceptor or donor depending on the underlying conditions. Among the defects reported in theoretical works are carbon substituting the gallium site ($C_{Ga}$), different configurations of interstitial carbon ($C_i$), carbon pairs (such as $C_{Ga}$-$C_N$ or $C_N$-$C_i$), as well as various carbon-containing complexes with intrinsic defects or other impurities such as hydrogen, oxygen, or silicon. Thus, the ability of carbon to form stable bonds with Ga, N and other C atoms results in a huge variety of possible stable point defects.[10,14–16,25] Some of these defects can play the role of donors and compensate or passivate acceptor doping. Hence, the identification of the most common types of carbon-containing defects and their properties is of paramount importance, and a critical step in understanding the electrical properties of carbon-doped GaN.[10,13–15,26] Moreover, as we will show, a high defect formation energy does not necessarily contradict a high density of such defects.

In this work, we demonstrate the identification of various carbon-related defects by combining methods of vibrational spectroscopy with calculations of the defect structure, vibrational frequencies, and atomic displacements using density functional theory (DFT). Comprehensive IR absorption and Raman studies of GaN samples doped with different carbon isotopes (GaN:C) are conducted to identify the spectroscopic fingerprints of the different carbon-based point defects in GaN. A series of eight $^{12}$C-doped GaN samples with varying doping concentration [C] = [$^{12}$C] + [$^{13}$C] between 3.2×10$^{17}$ and 3.5×10$^{19}$ cm$^{-3}$ and natural isotope $^{13}$C abundance 1.1% were grown to investigate the vibrational frequencies of the carbon point defects. In two additional samples with elevated concentration of the $^{13}$C isotope we observe a change of the vibrational frequencies of several peaks caused by C-doping. The analysis of the isotopic splitting via the harmonic oscillator approximation enables the identification of one, two, or three carbon atoms in the corresponding defect structure. Independent computation by DFT specified the structure of defects which could vibrate at these frequencies and clarified possible vibrational modes. In particular, we identify the origin of the recently observed vibrational modes at 765 cm$^{-1}$ and 774 cm$^{-1}$.[11,27,28] We unravel a correlation with two additional vibrational modes related to the same defect center and present a defect model explaining all four vibrational modes and their $^{13}$C-related replicas. Moreover, based on first-principles calculations we establish the structure of tri-carbon defects responsible for the intense IR absorption and calculate its charge states in the bandgap.[22,23] Finally, we consider several hitherto unnoticed peaks of IR absorption (carbon-hydrogen and carbon pairs) and demonstrate changes in the IR absorption intensity of most defect types under additional excitation at photon energies below the bandgap energy (2.71 eV and 3.22 eV). We also demonstrate methods to recognize the appearance of both $C_N$ and $C_N=C=C_N$ at carbon doping concentrations as low as $3.2 \times 10^{17}$ cm$^{-3}$.

## 2. Materials and Methods

Ga-polar c-plane oriented GaN layers, undoped or doped by carbon with natural or varied isotope composition were grown by hydride vapor phase epitaxy (HVPE) on 2 inch (0001) GaN/sapphire templates. The growth and doping procedure is described in detail in Ref. [18] The sources of carbon doping were: liquid pentane [Dockweiler Chemicals, electronic grade] or gaseous butane [Sigma Aldrich, Butane 12C4 (Gas) 99%] containing carbon in the natural isotopic composition, and gaseous butane isotopically enriched with $^{13}$C to 99% [Sigma Aldrich, Butane 13C4 (Gas) 99%], which was used to achieve carbon doping with 99% and 50% fraction of the isotope $^{13}$C. All layers separated from the sapphire substrates spontaneously after cool-down due to the large difference of thermal expansion coefficients. The c-plane oriented samples, with areas of about $10 \times 5$ mm$^2$ and $0.5 - 1$ mm thickness, were diced and both c-faces were polished. From two samples (GaN140 and GaN051-50; sample descriptions are provided below) cross-sectional m-plane stripes of about $5 \times 1 \times 0.5$ mm$^3$ were cut and polished. Concerning the crystalline perfection of the samples, we refer to the detailed investigation reported recently.[18] In-plane and out-of-plane strain of the c-plane GaN layers are negligible (the accuracy of the measurement of the lattice constants by X-ray diffraction was ±0.03% and ±0.01%, respectively), and the full width at half maximum of X-ray diffraction rocking curves at the symmetric (002) reflection and the skew symmetric (302) reflection vary between 80 and 120 arcsec. The results of these measurements are essentially independent of whether the layers are carbon doped or not, indicating that carbon doping does not impair the crystalline quality. The positions of the strain-related intrinsic Raman scattering bands only insignificantly differ and do not correlate with [C].

The chemical concentrations of carbon, oxygen, silicon, and hydrogen in the GaN layers were determined by secondary ion mass spectrometry (SIMS; performed by RTG Mikroanalyse GmbH Berlin). The numbers in the sample name indicate the value of [C] in units of $10^{17}$ cm$^{-3}$ (as given in Table I). In the undoped reference sample, [C], [O], and [Si] are below the respective SIMS detection limits: [C] $< 2.4 \times 10^{16}$ cm$^{-3}$, [O] $< 2 \times 10^{16}$ cm$^{-3}$, and [Si] $< 7 \times 10^{15}$ cm$^{-3}$. In the samples doped from pentane, [O] and [Si] are slightly above the detection limits of $3.5 \times 10^{16}$ cm$^{-3}$ and $1.3 \times 10^{16}$ cm$^{-3}$, respectively, while in the samples doped from isotopically enriched butane, [O] reaches $3.5 \times 10^{17}$ cm$^{-3}$. [H] is at about $1 \times 10^{17}$ cm$^{-3}$ for undoped GaN layers and does not exceed $8 \times 10^{17}$ cm$^{-3}$ for any carbon doping level in this study. Two samples, GaN051-50 and GaN025-99, have been doped with a significant fraction of $^{13}$C; the fraction (50%, 99%) is denoted as the second number in the sample name. The total carbon concentration [C] and the respective $^{13}$C percentage of each sample [$^{13}$C] / [C] are specified in Table I. Comparing [C] with the residual impurity concentrations mentioned above, [O] and [Si] attain at most one tenth of [C], while [H] never exceeds one third. Hence, compensation or passivation of carbon acceptors by these impurities is not expected to play a major role in this study.

Table I. Chemical concentrations of carbon isotopes and hydrogen in samples obtained by SIMS. * These are the samples denoted as C1213 and C13, respectively, in Ref.[22]

| Sample name | [C]=[$^{12}$C] + [$^{13}$C] $10^{17}$ cm$^{-3}$ | [$^{13}$C]/[C] | [H] $10^{17}$ cm$^{-3}$ | Precursor |
|---|---|---|---|---|
| GaN000 | < 0.24 | 1% | 1 | - |
| GaN003 | 3.2 | 1% | | Pentane |
| GaN005 | 5.0 | 1% | | Pentane |
| GaN019 | 19 | 1% | | Pentane |
| GaN058 | 58 | 1% | | Pentane |
| GaN090 | 90 | 1% | 6 | Pentane |
| GaN140 | 140 | 1% | | Pentane |
| GaN300 | 300 | 1% | | Pentane |
| GaN350 | 350 | 1% | 8 | Pentane |
| GaN051-50* | 51 | 50% | | Butane |
| GaN025-99* | 25 | 99% | | Butane |

Fourier-transform infrared (FTIR) absorption measurements were performed in the mid-infrared spectral range on a Bruker Vertex 80v spectrometer equipped with a globar source, a potassium bromide beam splitter, and a liquid nitrogen cooled mercury cadmium telluride detector at different temperatures. For measurements below RT, a liquid-helium-flow cryostat (Oxford OptistatCF) with zinc selenide windows was used. The spectra were recorded with a resolution of 0.5 cm$^{-1}$ at room temperature (RT) and 0.25 cm$^{-1}$ at 10 K. For polarization-dependent measurements, a holographic wire grid polarizer on a KRS-5 substrate was used. Additional FTIR measurements were carried out under below-bandgap illumination. The exciting light of wavelength 385 nm (3.22 eV) or 455 nm (2.71 eV) was guided by an optical fiber from a power light-emitting-diode source (Omicron LedHUB) into the sample chamber of the FTIR spectrometer using a vacuum-tight feedthrough. The fiber output was positioned in such a way that it was outside of the FTIR sample beam and that the LED divergent beam was incident under ~60° to the sample surface. Nominal LED power below 3000 mW was used in a continuous mode. Additional measurements in the range of Reststrahlen band absorption were done on sample GaN090 with a Bruker Tensor 27 spectrometer and an Oxford CF 2102 cryostat with KRS5 windows at liquid nitrogen temperature (Figure 1).

Raman experiments were carried out at room temperature (RT), with incident laser light at excitation wavelengths of 2.81 eV and 2.41 eV (442 and 514 nm) using a Horiba LabRAM HR800 Raman microscope spectrometer. Half-wave plates suited for the visible spectral range were used to change the direction of the electric field vector E of the incident, linearly polarized light from the exciting laser. The outgoing backscattered light was guided through a linear polarizer (analyzer). The orientation of the light beam and the polarization directions in the coordinate system of the GaN crystal (z∥c, x⊥c, y⊥c, x⊥y) are defined using the Porto notation: k($e_1 e_2$)k̲, where k and k̲ are the antiparallel directions of incoming and outgoing scattered light propagation, while $e_1$ and $e_2$ determine the orientation of electric field vector $\vec{E}$ of incoming and scattered light, respectively.

The methodology for DFT calculations is the same as in [10], which is outlined in more detail in [29]. Projector-augmented wave (PAW) pseudopotentials[30] are used within the VASP code[31], and the HSE hybrid functional[32,33] is employed to provide a correct description of the band gap and defect properties. 96-atom supercells are used with a 400 eV cutoff. Defect LVMs are calculated using the finite-difference method within VASP, and isotope effects are explored by adjusting atomic masses before recalculating the Hessian matrix. Two shells of atoms are included in the LVM calculations (i.e., nearest neighbors as well as next-nearest neighbors); as will be seen below this is important for obtaining relevant modes for $C_N$.

## 3. Results and Discussion

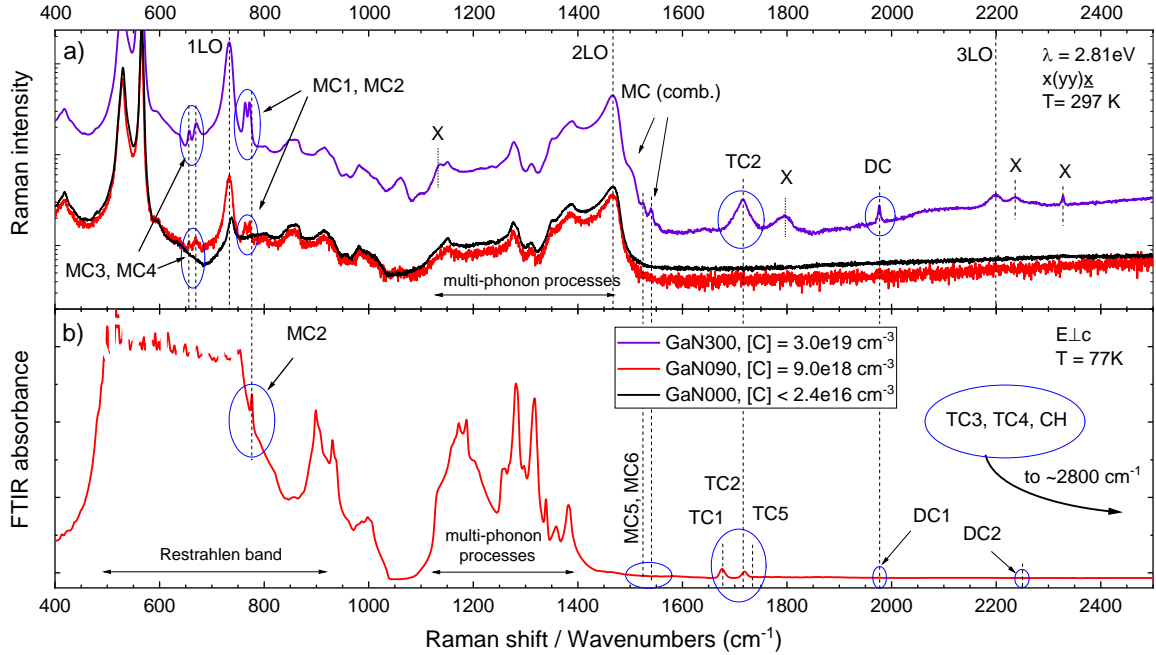

*Figure 1. Overview of carbon defect vibrational modes in GaN:C. (a) The room temperature Raman spectra are shown for the reference carbon-free sample ([C]<2.4×10$^{16}$ cm$^{-3}$) and two C-doped samples GaN090 and GaN300 at 2.81 eV excitation energy and x(yy)x̄ geometry. (b) The absorption spectrum of sample GaN090 is taken at E⊥c polarization and a temperature of 77 K. Peaks labeled: MC (mono-carbon), DC (di-carbon), and TC (tri-carbon) are related to carbon-defects of different groups, LO indicates longitudinal phonon modes, and X indicates peaks of unclear origin. The peaks TC3, TC4, and CH (carbon-hydrogen) in the range of 2670 – 2850 cm$^{-1}$ are shown in Fig. 5 (b).*

The broad-range spectra of Raman scattering at room temperature (a) and IR absorption at 77 K (b) are shown in Figure 1. Raman spectra in Figure 1(a) of the undoped reference sample (black solid line) and two different C-doped GaN crystals reproduce similar general spectral features originating from intrinsic vibrations. The peaks at 532 cm$^{-1}$ and 567 cm$^{-1}$ are identified as intrinsic vibrational modes $A_1(TO)$ and $E_2(high)$ of the GaN lattice allowed for the selected polarization geometry x(yy)x̄.[34–36] The mode $E_1(LO)$ at 741 cm$^{-1}$ has the highest frequency within the first-order phonon spectrum of GaN thus limiting the range of *intrinsic* first order Raman modes (labeled 1LO). Naturally, the 2LO and 3LO frequencies define the ranges of second and third order intrinsic vibrations, respectively. Multi-phonon processes are clearly visible in the second-order range between 1LO and 2LO. Accordingly, the IR absorption spectrum of sample GaN090 (Figure 1 (b)) contains signals in the range 1100 – 1500 cm$^{-1}$ caused by two-phonon processes, while the opacity at lower wavenumbers corresponds to the Restrahlen band.[37]

Apart from the first and higher order host lattice vibrations, additional peaks of carbon related defect vibrational modes appear in the C doped spectra in Figure 1. They are marked by blue circles. Labels MC, DC, and TC group the peaks for further discussion, where individual peaks are labeled with additional numbering (TC1, DC2, etc.). The exact peak positions are summarized in Table IV at the end of the paper. Peaks labeled with X are of unknown origin and not further discussed within this work.

To unravel the defect structure related to the observed vibrational signatures, we investigate the isotope effect in carbon doped samples with controlled $^{12}$C and $^{13}$C doping and analyze the frequency shifts within the harmonic oscillator approximation. The vibrational frequency $v$ of the oscillator given in wavenumbers (cm$^{-1}$) is determined by the masses of vibrating atoms ($m_i$) and the stiffnesses of the chemical bonds. These two parameters also determine the amplitude of vibration. The replacement of a carbon atom by another isotope preserves[1] the chemical stiffness (and hence the oscillator energy) but provides a change of $v$ and vibrational amplitudes of the system due to the change of atomic mass ($m \rightarrow m^*$). The relative frequency shift is expressed by frequency ratio $f \equiv \frac{v}{v^*}$.

It should be noted that the impact of the individual atom to the change of vibrational frequency is proportional to the amplitude of its vibration squared. In the special case of a vibration involving only two mobile atoms ($i=2$), replacing $m_1, m_2 \rightarrow m_1^*, m_2^*$ results in the frequency ratio:

$$\frac{v}{v^*} = \left(\frac{m_1^* m_2^*(m_1 + m_2)}{m_1 m_2 (m_1^* + m_2^*)}\right)^{0.5} \quad (1)$$

When both vibrating atoms are $^{12}$C and replaced by $^{13}$C, the frequency ratio reaches the theoretical maximum $f_{max} \approx 1.041$ for carbon atoms. Note that this is a constant value, not only for the simplest case of the diatomic oscillator, but also for any vibrating structure when all $n$ carbon atoms with nonzero vibrational amplitude are replaced and there are no other vibrating atoms. For example, $f_{max}$ was demonstrated earlier for tri-carbon defects in AlN and GaN.[22,38] The mobile but unreplaced atoms result in frequency ratio $f$ significantly smaller than $f_{max}$ as one can observe, e.g., for antisymmetric stretching modes of linear C-C-N and N-C-N defects in GaAs.[39,40] Furthermore, host atoms around the defect, being involved in the vibrational mode, impact the vibrational mode in the same manner decreasing $f$. In the present work, we use DFT to evaluate the impact of host atoms on the value of the frequency ratio $f$ that becomes critically important for certain vibrational modes of $C_N$ acceptors, because these modes exhibit host-atom involvement.

The number of carbon atoms in the defect structure is determined by the splitting of the original vibrational mode frequency in the sample with equal concentrations of two isotopes (GaN051-50) due to the maximum diversity of isotopomers[2] in this sample. At [$^{12}$C] = [$^{13}$C] = 50% all $N$ isotopomers become statistically equivalent, and hence, they are present in equal quantity in the vibrational spectra. Then the number of isotopomers $N = 2^n$ is equal to the number of combinations of *two* isotopes on $n$ sites of carbon atoms in the defect structure. The number of unique frequencies in this case is equal to or lesser than $2^n$, taking into account degeneracies of structurally equivalent isotopomers of symmetrical defects.[22]

More information about the structure can be obtained from the measurements of IR absorption polarization. The IR absorption of the vibrating dipole $\Delta\vec{\mu}$ of a defect vibrational mode in a crystal is maximum when $\Delta\vec{\mu} \parallel \vec{E}$ and zero at $\Delta\vec{\mu} \perp \vec{E}$. The absorption $A(\varphi,\gamma)$ as a function of the angles $\gamma$ (the tilt angle between $\Delta\vec{\mu}$ and wurtzite symmetry axis c) and $\varphi$ (between $\vec{E}$ and c) can be expressed by the equation, taking into account the symmetry properties of GaN crystals:[41]

---

[1] at least for carbon and heavier atoms the change is insignificant.

[2] i.e., defects different only in the isotope composition. One can distinguish three groups of isotopomers: "$^{12}$C-pure" and "$^{13}$C-pure" containing only $^{12}$C or $^{13}$C atoms, respectively, and "mix" isotopomers containing both isotopes.

$$A(\varphi,\gamma) = A_0(sin^2\gamma + (2 - 3sin^2\gamma) \cdot cos^2\varphi) \quad (2)$$

where $A_0$ is the normalization constant. From this equation the orientation of a defect can be found just by two measurements of the linearly polarized absorption coefficient at E∥c ($\varphi=0°$) and E⊥c ($\varphi = 90°$):

$$P = \frac{A(90,\gamma)}{A(0,\gamma)}, \gamma = arcsin\left(\frac{2}{2+P}\right)^{1/2} \quad (3)$$

Following the previous discussion on the general aspects of the isotope effect in vibrational spectroscopy of carbon doped GaN, the subsequent sections focus on the identification of specific mono-carbon (MC), di-carbon (DC) and tri-carbon (TC) defect complexes.

### a. Mono-C defects and CH modes

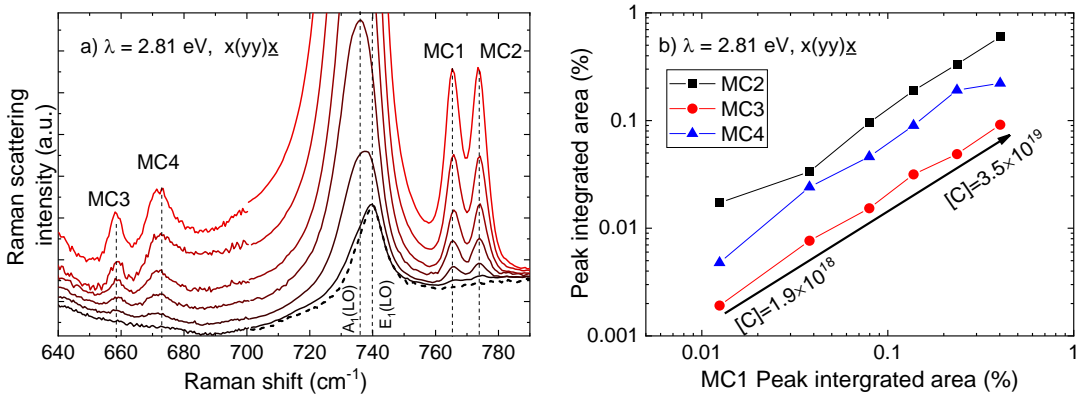

*Figure 2. (a) Raman scattering spectra of samples at [C] in the range from $5\times10^{17}$ cm$^{-3}$ to $3\times10^{19}$ cm$^{-3}$ measured on the m-facet of c-plane wafers in the x(yy)x̄ geometry at room temperatures. The individual spectral curves are normalized to the integrated area in the range from 150 to 1535 cm$^{-1}$. (b) the corresponding integrated peak area for vibrational modes MC1 – MC4 for the samples at [C]=$1.9\times10^{18}$ cm$^{-3}$ and above.*

**Vibrations of $C_N$ defect.** Figure 2 shows the Raman spectra of the reference GaN (dashed) and 6 GaN:C samples with different carbon concentration between $5\times10^{17}$ cm$^{-3}$ to $3\times10^{19}$ cm$^{-3}$ in the spectral range between 640 and 790 cm$^{-1}$. Since the Reststrahlen band blocks IR transmission between 600 – 800 cm$^{-1}$ (Figure 1 (b)), Raman scattering is the preferable method for investigating defect vibrations in this range. Apart from the dominating host GaN lattice modes $A_1$(LO) and $E_1$(LO), 4 well separated peaks can be identified whose intensity increases with increasing carbon doping. These 4 scattering peaks MC1 – MC4 (see Table II) are recognized vibrations of carbon-based defects. They become very intense at high [C] and have proportional integrated area as it is shown in Figure 2 (b) for the most accurate measurements (at [C] above $1.9\times10^{18}$ cm$^{-3}$). Scattering peaks with frequencies close to true MC3 is often observed independently due to doping with various impurities; one should, however, interpret the data carefully in this case. As will be shown below, the nature of the MC3 vibration is highly determined by the GaN lattice, while the influence of the impurity atom on the vibration frequency is of secondary importance. In this respect, the observation of similar vibration frequencies (655 – 670 cm$^{-1}$)[42–45] for Mn, Mg and different implanted atoms should not confuse. The group of 4 always proportional scattering peaks MC1 - MC4 provides reliable agreement with the model of only one type of substitutional C defect computed by DFT.

*Table II. The vibrational frequencies found from the Raman spectra and calculated via DFT for different models of defects $C_N$, $C_{Ga}$, and $C_i$.*

| Experiment | | DFT | | | | | |
|---|---|---|---|---|---|---|---|
| $^{12}C$, cm$^{-1}$ | $f$ | $^{12}C_N^-$, cm$^{-1}$ | $f$ | $^{12}C_{Ga}^+$, cm$^{-1}$ | $f$ | $^{12}C_i^0$, cm$^{-1}$ | $f$ |
| 766 (MC1) | 1.028 | 775 | 1.030 | 843 | 1.027 | 1511 | 1.021 |
| 774 (MC2) | 1.028 | 785 | 1.030 | 836 | 1.026 | 734 | 1.027 |
| 659 (MC3) | 1.000 | 655 | 1.00 | 831 | 1.026 | 727 | 1.009 |
| 672 (MC4) | 1.000 | 662 | 1.00 | 735 | 1.000 | 466 | 1.015 |

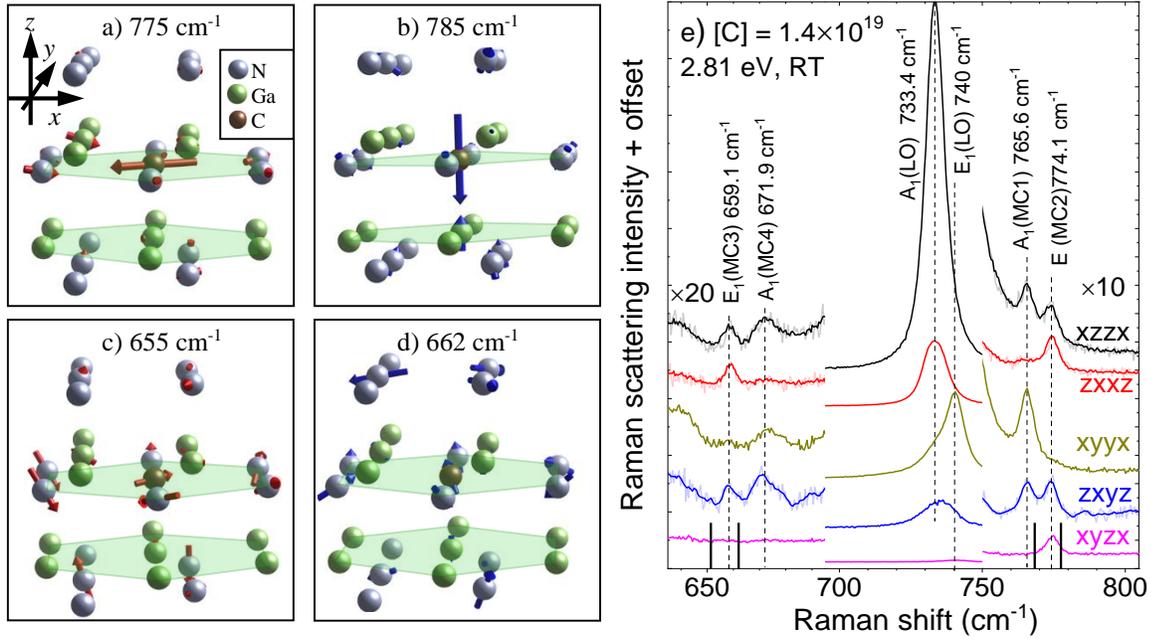

*Figure 3. (a – d) Displacements of atoms (shown by arrows) according to DFT calculations in the four vibrational modes of the $C_N$ defect are found at vibrational frequencies 775, 785, 655, and 662 cm$^{-1}$. The frequencies can be associated with modes MC1, MC2, MC3, MC4, respectively, presented in Raman spectra (e) by stick spectra.*

Within the calculated (most expected) defects with only one carbon atom in the structure ($C_{Ga}$, $C_N$, $C_i$ in Table II) only $C_N$ in the negative charge state ($C_N^-$) (Figure 3 (a – d)) possesses 4 vibrational frequencies close to the optical phonon band and can perfectly explain all 4 peaks MC1

– MC4 simultaneously including their polarization properties. In contrast, $C_{Ga}$ and $C_i$ defects possess too different vibrational frequencies to associate them with the peaks MC1 – MC4. A mode near 785 cm$^{-1}$ as well as a doubly degenerate mode near 775 cm$^{-1}$ were found by DFT previously.[10] As shown in Figure 3 (a, b), these modes are dominated by displacements of the C impurity atom. Further two modes are found for $C_N^-$ are doubly degenerate modes at 662 and 655 cm$^{-1}$. These modes were not observed in earlier reports, as the nearest in-phase nitrogen shells were not used to construct the Hessian matrix.[10,11] The character of these modes can be seen in Figure 3 (c, d). Unlike the highest-wavenumber modes, these modes have only weak carbon character due to many in-phase vibrating nitrogen atoms located around $C_N$. The type of vibrations clarifies the irreducible representations of these vibrational modes and hence polarization selection rules. In Figure 3 (e) there are the spectral curves of sample GaN140 for 5 different geometries captured from the facet (i.e., geometries x(zz)x, x(zy)x, x(yy)x) or from the c-plane (i.e., geometries z(xx)z and z(xy)z). In this respect, explaining modes MC1 – MC4 we exclude the other defects with one carbon atom. Utilizing the isotope effect, we show below that defects with more than one carbon atom can also be excluded.

Table II represents the isotopic shifts of vibrational frequencies (as frequency ratio *f*) due to substituting $^{13}$C for $^{12}$C in DFT calculations for three different types of mono-carbon defects, $C_N$, $C_{Ga}$, and $C_i$. MC3 and MC4 vibrational modes belong to the range of frequencies propagating in GaN crystals (the vibrational band formed by the optical phonon branches). The specifics of this case, also known as the band mode, is that the major part of the vibrational energy is stored in the vibrational displacement of many host atoms of the crystal. Thus, a change in the mass of the impurity atom during isotopic replacement does not affect the oscillation frequency. Modes 662 cm$^{-1}$ and 655 cm$^{-1}$ satisfy this condition and DFT calculations provides less than 1 cm$^{-1}$ shift for the modes (*f*=1.00) for the supercells considered here. MC1 and MC2 modes, having vibrational frequencies near to the range of allowed vibrations, are also influenced by the host atoms. We find that the frequency shift to lower wavenumbers depends on the number of the host atoms considered in the model. An increase in the number of gallium and nitrogen atoms reduces the relative impact of the carbon atom leading to a decrease of *f* from 1.036 (1 carbon and 3 neighboring Ga atoms) to 1.030 (3 Ga-N shells around C atom). In contrast, changing the number of neighboring atoms gives no visible effect for the localized vibrational modes of $C_{Ga}$ and $C_i$.

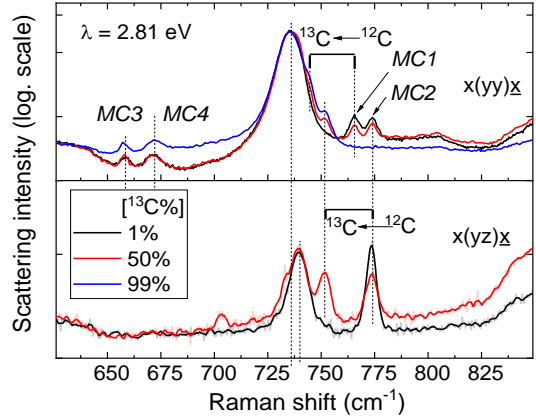

*Figure 4. Raman scattering spectra of samples with different isotope abundance [$^{13}$C%] equal to 1% (GaN019), 50% (GaN051-50), and 99% (GaN025-99) at 2.81 eV excitation for two different geometries of scattering, x(yy)x and x(yz)x, available from the facet of the c-plane wafers (normalized to the intrinsic LO peak height).*

Isotopic shifts of $f = 1.00$ (MC3, MC4) and $f = 1.028$ (MC1, MC2) are demonstrated in Figure 4. The vibrational frequencies of the MC1 and MC2 modes associated with $^{12}$C isotopes (black line) are equal to 766 and 774 cm$^{-1}$. Geometries x(yy)x and x(yz)x allow to distinguish between these two modes. Accordingly, the two peaks with similar polarization properties appear in both samples with an elevated abundance of $^{13}$C isotopes at a lower frequency of vibration (745 and 753 cm$^{-1}$, respectively), resulting in $f = 1.028$ for both the MC1 and MC2 modes. The $^{13}$C-related pair of peaks is exclusively observed in samples with significant concentrations of $^{13}$C, and associated with the $^{13}$C pure isotopomers of the corresponding defects It is also important to note that for the GaN051-50 sample, there are no additional peaks between the $^{13}$C and $^{12}$C peaks that would be associated with the mixed isotopomers of possible multicarbon defects. Thus, the double splitting of each of the MC1 and MC2 modes indicates the involvement of a defect with one C atom. The frequencies of the band modes MC3 and MC4 do not change in $^{13}$C enriched samples for the reasons mentioned above. Since the number of vibrating host atoms is much larger in the real crystal, the experimental value $f = 1.028$ for modes MC1 and MC2 is slightly smaller than the theoretical value, as expected.

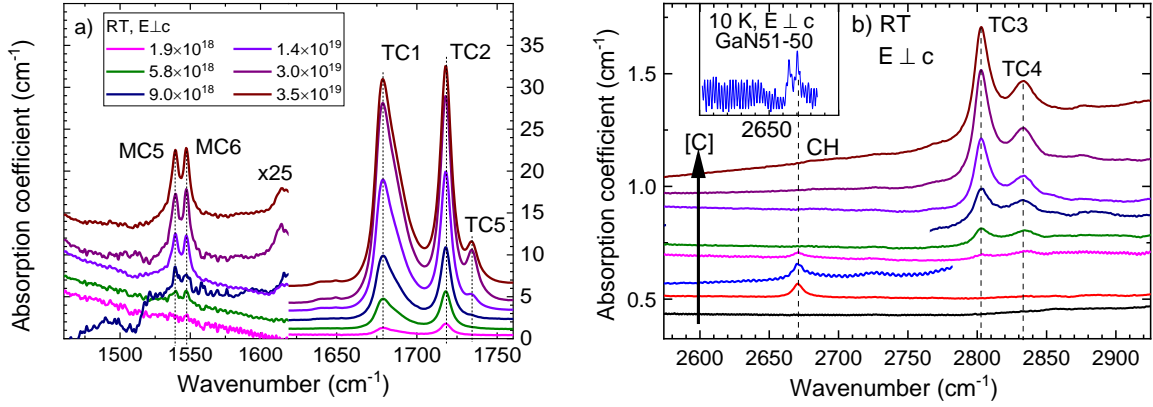

*Figure 5. (a) IR absorption spectra of samples with different [C] containing IR absorption peaks of mono-C defects (MC5 and MC6) after the subtraction of the spectral curve of sample GaN005. Modes TC1, TC2 are shown for intensity comparison. (b) IR absorption spectra of all 9 GaN samples at room temperature and polarization E⊥c in the range of high wavenumbers. Peaks MC5, MC6, CH, TC1 – TC5 are marked by vertical drop lines. The insert illustrates the double-splitting peak CH measured at 10 K for the same polarization in sample GaN051-50 with equal concentrations of $^{12}C$ and $^{13}C$ isotopes. See more in Fig. 10 (c).*

We report also two new faint vibrational modes MC5 and MC6 appearing in the IR absorption spectra at 1539 and 1547 cm$^{-1}$ at RT (1544 and 1553 cm$^{-1}$ at 10 K) for E⊥c polarization. The $^{13}C$-related modes are observed at 1501 and 1510 cm$^{-1}$ in both samples GaN025-99 and GaN051-50 also without additional peaks of mixed isotopomers in the sample GaN051-50. We note that though a mode exactly equal to 1539 cm$^{-1}$ was predicted previously for $C_i^0$,[10] nevertheless, DFT calculations indicate that the isotopic effect for $C_i$ mode is expected to be $f = 1.021$, significantly lower than the experimental value, $f = 1.0285 \pm 0.0005$ (the same as for MC1 and MC2).

The exact match of $f$ allows us to identify the peaks MC5 and MC6 as the second-order harmonic of the MC2 vibration observed in the IR absorption spectra at E⊥c at 774 cm$^{-1}$ (Figure 1 (b)). Mode MC1 does not exhibit at E⊥c polarization hence the second order signals is not expected too (in our measurements of IR absorption, only faint MC1 peaks were observed at E∥c). The double-headed structure of the second-order signal is not clarified in this work. One possible explanation could be a Fermi-resonance effect, since the original MC2 mode is doubly degenerate.

**CH vibrations.** In addition to substitutional mono-carbon on GaN host lattice sites, carbon-hydrogen vibrations can occur which are commonly observed in various semiconducting materials at high frequencies above 2600 cm$^{-1}$.[46,47] Figure 5 (b) shows the peak at 2670.3 cm$^{-1}$ labeled "CH" present in samples with [C] = 1.9×10$^{18}$ cm$^{-3}$ and lower near the TC3 and TC4 modes that in turn appear at higher [C]. In contrast to peaks TC3 and TC4, the CH peak splits into only 2 peaks in the spectra of sample GaN051-50 (see insert of Fig. 5(b)). The peaks in this pair are close to each other ($f = 1.002$). Both peaks have the same intensity, and since there are no more peaks with that intensity in this range, only a single carbon atom is expected to be a constituent of the defect structure.

According to Eq. 1, the extremely small frequency shift (~6.1 cm$^{-1}$) points to a neighbor of C being a very light atom (i.e., hydrogen). With $m_1 = 12, m_2 = m_2^* = m_H = 1, m_1^* = 13\ a.m.u.$, $v_{CH} = 2670.3\ cm^{-1}$ Eq. 1 gives $v_{CH}^*$ equal to 2662.7 cm$^{-1}$, close to the experimental value (2664.2 cm$^{-1}$). The difference can be explained mostly by the deviation of the oscillator from the harmonic model. The result seems to be acceptable in comparison with the data for other semiconductors containing gallium, e.g. GaAs and GaP.[46,47] The vibrational modes for $^{12}$C-H ($^{13}$C-H) observed at 2660.2 (2652.6) cm$^{-1}$ and 2635.1 (2628.4) cm$^{-1}$ have a mismatch within the harmonic model of 0.3 cm$^{-1}$ and 1.25 cm$^{-1}$, respectively.[46,47] In this way, the observations of CH isotopic shifts and the absolute frequency range are in good agreement with stretching vibrations of C-H bonds.

### *b. Di-C defects*

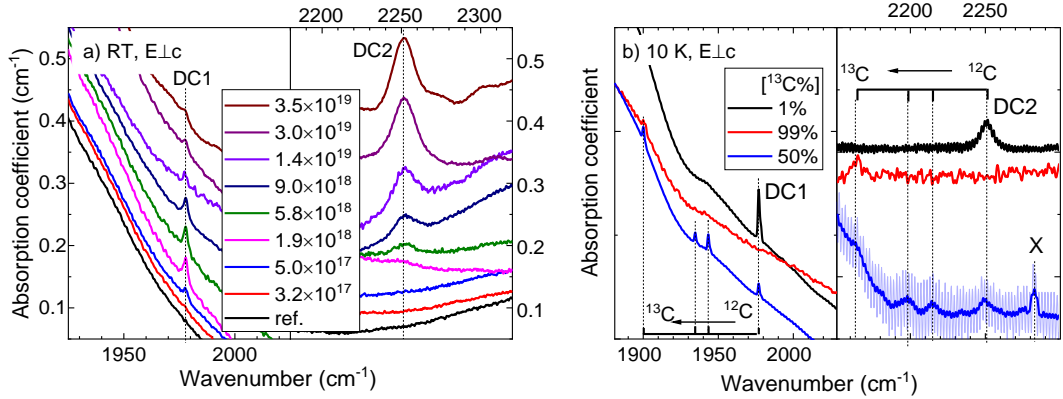

*Figure 6. (a) Uncorrected IR absorption spectra in the range of DC1 and DC2 modes for the reference GaN and all 8 GaN:C samples with natural isotope composition at room temperature (RT) for E⊥c polarization. (b) the modes DC1 and DC2 at different isotope abundance: natural [$^{13}$C%] ≈ 1% (GaN058), [$^{13}$C%] ≈ 50% (GaN051-50), and [$^{13}$C%] ≈ 99% (GaN025-99) measured at 10 K for the same polarization. All spectra are shifted vertically for visualization.*

Following the discussion of mono-carbon and C-H vibrations, we now focus on the identification of di-carbon defects, in particular, $C_i$-$C_N$ and C≡C defects. The carbon-pair defects are present in most samples, but only as weak signals in the FTIR (DC1, DC2) and Raman (DC1, not shown) spectra. Figure 6 (a) illustrates the intensity of the mode at room temperature and E⊥c polarization for all 8 GaN:C samples. The isotope effect in crystals with carbon isotope abundances [$^{13}$C] equal to 1% (GaN058), 50% (GaN051-50) and 99% (GaN025-99) is clearly visible in Figure 6 (b) for measurements at 10 K and E⊥c polarization. The similar splitting of both DC1 and DC2 modes into four individual frequencies in sample GaN051-50 indicates that two carbon atoms are present in the defect structure at two inequivalent structural positions (XY-defect). In the spectra of both $^{13}$C-enriched samples, $^{13}$C-pure isotopomers provide vibrations at 1900.2 and 2164.2 cm$^{-1}$. The frequency ratio then corresponds precisely to $f_{max} \approx 1.040 \pm 0.0005$ given by the isotope masses.[22]

The range of 2000–2300 cm$^{-1}$ corresponds to stretching vibrations of C=N, C≡N, C=C, and C≡C pairs and antisymmetric stretching vibrations of carbon and nitrogen atoms.[48] We exclude any vibrations involving non-carbon atoms (e.g., antisymmetric stretching of C=N=C or C=C=N defects) based on the observations of $f_{max}$ in our cases.[10,14] According to earlier DFT calculations of the $C_i$-$C_N$ defect, the carbon pair tilts by 55.8° relative to the c-axis, providing a similar frequency of vibration (2049 cm$^{-1}$ or 2057 cm$^{-1}$) when in the "+" charge state,[10,14] in good agreement with DC1. The dependence on polarization shown in Figure 7 provides further evidence. According to Eq. (3), the absorption peak areas of the DC1 and DC2 peaks measured at two polarizations E⊥c and E∥c provides the values of the tilt angles as approximately 55° and 30°. The DC2 mode does not fit to any results of the calculations, and thus cannot yet be assigned.

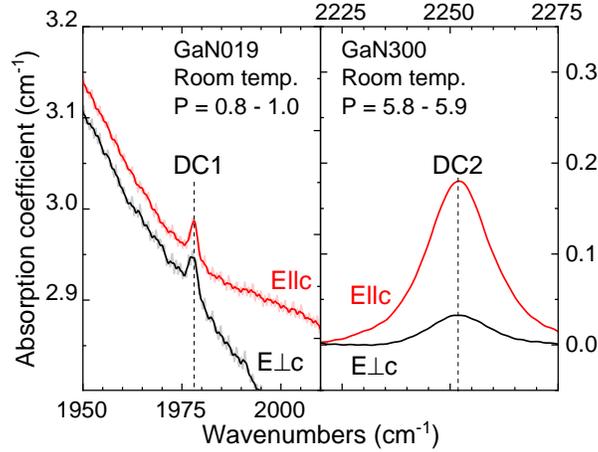

*Figure 7. IR spectra of samples GaN019 (DC1, smoothed) and GaN300 (DC2) for two polarizations E ⊥ c and E ∥ c. Mode DC1 has low intensity nearly equal at both polarizations (P ≈ 0.8 – 1.0). Peak DC2 prevails at E ∥ c polarization (P ≈ 5.8 – 5.9). The tilt angle of corresponding defects can be evaluated via Eq.3*

### c. Tri-C defects

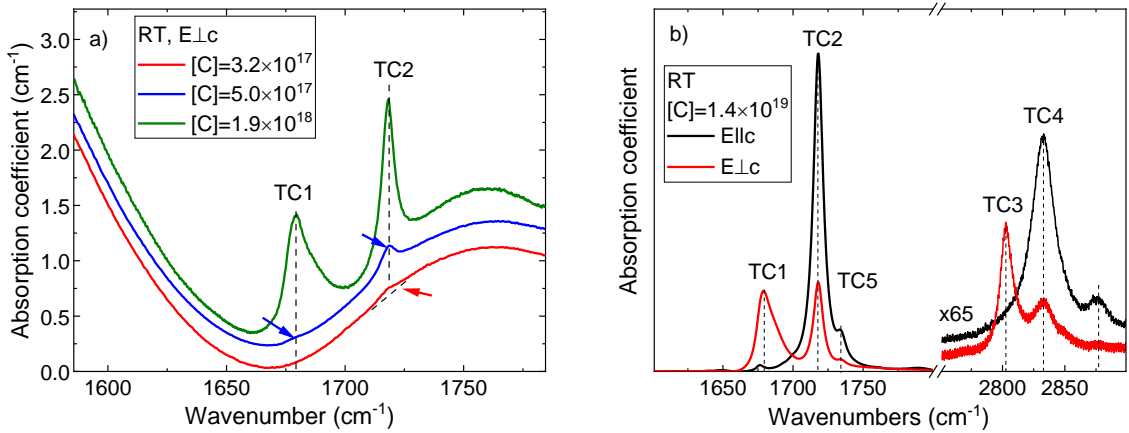

*Figure 8. (a) IR absorption spectra of samples GaN003, GaN005, GaN019 at room temperature (RT) and E⊥c polarization containing peaks TC1 and TC2. (b) Polarization dependence is shown for sample GaN140. Peaks TC2 and TC4 dominate for E∥c polarization but appear at both polarizations, while TC1 and TC3 are detectable at E⊥c polarization only.*

Finally, we turn to the most complex type of carbon-defect that is observed in GaN:C, i.e., the vibrations that involve three connected carbon atoms. These defects are labeled as tri-carbon (TC). Our previous works showed high prevalence of tri-C defects in GaN material. [18,22,23] Analysis of the isotope splitting and polarization dependence in GaN:C established the structure of defects covered behind the vibrational modes TC1 and TC2 derived exclusively from the analysis of vibrational spectra.[18,22,23] In this work, we consider this pair of signals together with their higher frequency satellites labeled TC3 and TC4, because similar effects are observed for these peaks. Furthermore, we propose a DFT model of tri-C defects explaining the frequencies of vibration with high accuracy. Mode TC5 appears only at the highest [C] and probably related to similar defects.

Figure 8 (a) shows that the TC1 and TC2 modes at 1673.9 and 1717.8 cm$^{-1}$ from tri-C defects are still visible for the lowest concentrations of carbon doping within the examined range in this work (GaN003 with [C] = 3.2×10$^{17}$ cm$^{-3}$, red line), hence, tri-C defects have manifested at this relatively low [C]. Figure 8 (b) displays the polarization dependence of the IR absorption for [C] = 1.4×10$^{19}$ cm$^{-3}$. Apparently, the spectra for E∥c polarization are dominated by TC2 and TC4 and TC1 and TC3 for the E⊥c polarization. Using Eq. (3), the observed polarization dependence directly shows that the dipole moment of defects vibrating in the TC1 mode is oriented in the basal plane (absorbing only at E⊥c polarization) while the dipole moment of the TC2 vibrational mode has a tilt angle of 33° relative to the GaN crystal c-axis (absorbing at both polarizations).[22,23]

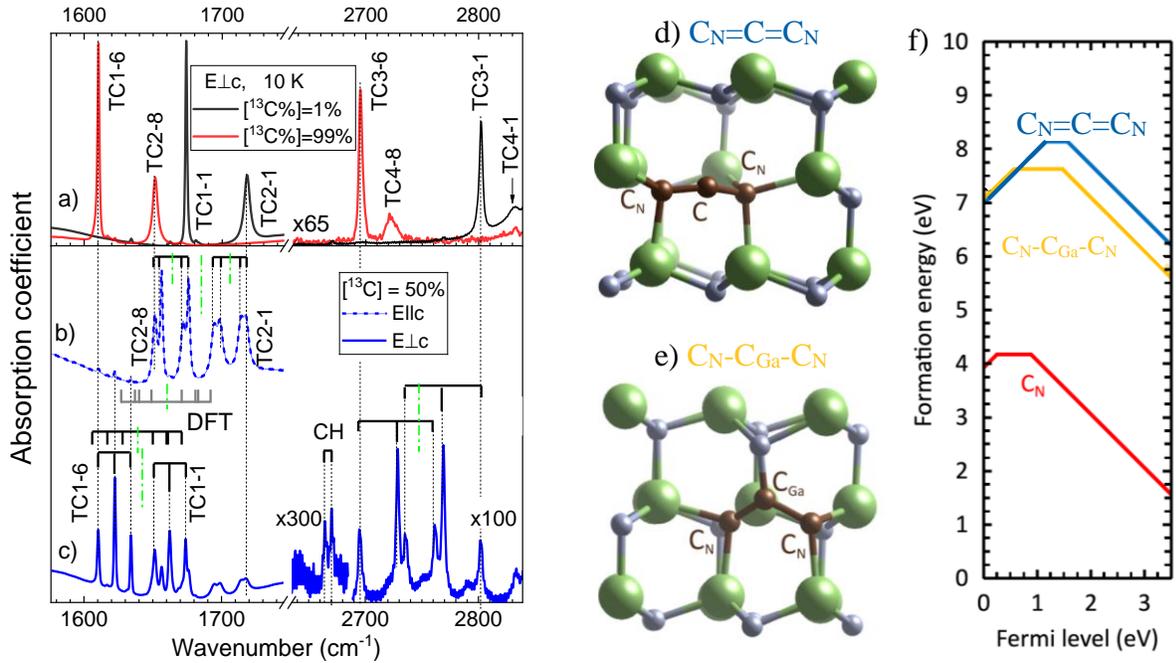

*Figure 9.(a) IR absorption spectra of samples with different isotope $^{13}C$ abundance [$^{13}C\%$]=1% (GaN019) and 99% (GaN025-99) for E⊥c polarization. (b, c) IR absorption spectra of sample GaN051-50 ([$^{13}C\%$]=50%) for E⊥c (b) and E∥c (c) polarizations. The 8-fold splitting of TC1 (b) and 6-fold splitting of TC2 (c) were discussed in [22]. The splitting of mode TC3 shown in (c) has identical nature and corresponds to the same triatomic structure. Splitting of the faint mode TC4 could not be observed. All spectra are captured at 10 K. (d) DFT model of basal tri-C defect found in this work vibrating at $v_3$ = 1671 cm$^{-1}$. (e) Configuration of basal tri-C defect reported earlier by Lyons et. al and vibrating at $v_3$ = 1152 cm$^{-1}$.[10] (f) Formation energy versus Fermi level for the two configurations of tri-C complexes compared with $C_N$..[10,13] Frequencies obtained for TC1 and TC2 via DFT are shown with the stick-spectra.*

The structure of complexes in previous theoretical reports [10,14–16] cannot explain the extensive data provided by vibrational spectroscopy (Figure 9 (a - c)). Using hybrid DFT calculations, the present work discovers new stable basal and axial configurations of tri-C complexes (the basal one is shown in Figure 9 (d)). For the two configurations (basal, axial) of tri-C defects, DFT provides the symmetric stretching ($v_1$) modes at (1165 cm$^{-1}$, 1192 cm$^{-1}$) and antisymmetric stretching ($v_3$) modes at (1671 cm$^{-1}$, 1691 cm$^{-1}$). As was established in Ref.[22], the TC2 mode is the $v_3$ mode of the bent $C_X$=$C_Y$=$C_Z$ triatomic chain of three C atoms (n = 3) at three unique structural positions (X, Y, Z) as confirmed by an 8-fold isotopic splitting. This 8-fold splitting (labeled TC2-8 to TC2-1) is clearly visible in GaN051-50 sample in Figure 9 (b). Performing $^{12}C \rightarrow ^{13}C$ isotopic replacement in hybrid DFT calculations (Table III) reproduces the 8-fold isotope splitting (in Figure 9 shown by the stick-spectra). The 6-fold splitting of TC1 (Figure 9 (c)) indicates a $C_X$=$C_Y$=$C_X$ structure, with symmetrical positions of the outer atoms (Z = X).[22]. In DFT, the basal configuration provides only 6 unique frequencies $v_3$ and 3 unique frequencies $v_1$ (Table III). While the 6-fold splitting emphasizes the symmetry of the basal defect structure,[22] the splitting of the $v_1$ mode into only 3 unique frequencies is assigned to additional degeneration caused by the symmetry of the *symmetrical* vibrational mode and the high values of the angle between the two carbon-carbon bonds. For these two reasons, the vibrational amplitude of the central C atom in the $v_1$ mode according to DFT is ~20 times smaller than for the outer C atoms explaining the degeneration to 3 unique $v_1$. To sum up, in all listed aspects the new DFT model satisfies experimental observations including $f \approx f_{max}$ for the complete isotope replacement.

The frequencies in the range of $v_3$ ~1700 cm$^{-1}$ and the equilibrium distance between the carbon atoms of 0.134 nm are typical for double carbon-carbon (C=C) bonds.[49–51] In contrast, the tri-C defect with the equilibrium distance 0.148 nm ($C_N$-$C_{Ga}$-$C_N$) reported earlier by Lyons et al. (Figure 9 (e)) shows a much lower frequency $v_3$ ~ 1152 cm$^{-1}$ and larger bond distances expected for single (C-C) bonds.[10,51] The difference in the bond order appears due to the breaking of the chemical bond between the central C atom and the neighboring nitrogen atom, together with a large deviation of the central C atom from the original position of the substituted Ga atom. The calculations provide an angle of 161° between the $C_N$=C bonds for the basal defects, versus 134±8° evaluated from the isotope splitting.[22] It is equivalent to the deviation of the central C atom from the position being substituted Ga atom slightly larger in the DFT model than it follows from the isotopic experiment. In this respect, the new configurations can be labeled as $C_N$=C=$C_N$, emphasizing the difference with configuration $C_N$-$C_{Ga}$-$C_N$ from the earlier report.

Despite the frequencies of the vibrational modes TC3 and TC4 lie in the typical range of carbon-hydrogen vibrations, the isotopic effects clearly exclude C-H as the origin of TC3 and TC4. The isotopic effects observed for TC3 and TC4 include the splitting to 6 individual frequencies with frequency ratio $f \approx 1.0400\pm0.0005$, identical to the value found for TC1 and TC2 and close to $f_{max}$. Furthermore, the 6-fold splitting to two symmetrical 1:2:1 triplets (Figure 9 (c)) for TC3 (similar to one observed for TC1) indicates $C_X$=$C_Y$=$C_X$ structure of the defect.[22] In this respect, TC3 and TC4 modes at 2802 and 2833 cm$^{-1}$ can be considered as combination vibration ($v_1 + v_3$) of $C_N$=C=$C_N$ defects with frequencies close to their algebraic sum of the basic modes $v_1$ and $v_3$. One can compare $C_N$=C=$C_N$ defects in GaN with the laser-ablated tri-C clusters vibrating at $v_3$ equal to 2040 or 1722 cm$^{-1}$ (for two different configurations).[49,50] The combination vibration ($v_1 + v_3$) of one of the configurations of such clusters are found experimentally at 3250 cm$^{-1}$ with anharmonicity below 10 cm$^{-1}$.[49,50] Then, the vibration frequency $v_1$ for the basal $C_N$=C=$C_N$ defect can be eval-uated as ~1130 cm$^{-1}$ (as the difference between TC1 and TC3 without taking into account the anharmonicity effect)[49]. Thus, our DFT model overestimates $v_1$ (1165 cm$^{-1}$, Table III) by ~1% which can be regarded as good agreement.

*Table III. Symmetrical ($v_1$), antisymmetrical ($v_3$) and combination ($v_1 + v_3$) vibrational mode frequencies of different isotopomers obtained from IR absorption spectra at 10K and from DFT.*

| Isotopomer | Frequency of basal Tri-C defect, cm$^{-1}$ | | | | | | | Frequency of axial Tri-C defect, cm$^{-1}$ | | | | | | |
|---|---|---|---|---|---|---|---|---|---|---|---|---|---|---|
| | $v_3$ | | | ($v_1 + v_3$) | | | $v_1$ | $v_3$ | | | ($v_1 + v_3$) | | | $v_1$ |
| | Peak TC1 | Exp. | DFT | Peak TC3 | Exp. | Exp.* | DFT | Peak TC2 | Exp. | DFT | Peak TC4 | Exp. | Exp.* | DFT |
| $^{12}$C-$^{12}$C-$^{12}$C | 1 | 1673.9 | 1671 | 1 | 2803.5 | ~1130 | 1165 | 1 | 1717.8 | 1692 | 1 | 2831 | ~1115 | 1182 |
| $^{13}$C-$^{12}$C-$^{12}$C | 2 | 1662.3 | 1661 | 2 | 2769.3 | ~1107 | 1142 | 2 | 1714.4 | 1683 | - | - | - | 1158 |
| $^{12}$C-$^{12}$C-$^{13}$C | | | 1660 | | | | 1142 | 3 | 1699.0 | 1681 | - | - | - | 1160 |
| $^{13}$C-$^{12}$C-$^{13}$C | 3 | 1651.5 | 1650 | 3 | 2736.7 | ~1085 | 1120 | 4 | 1694.6 | 1671 | - | - | - | 1137 |
| $^{12}$C-$^{13}$C-$^{12}$C | 4 | 1634.1 | 1628 | 4 | 2760.6 | ~1127 | 1164 | 5 | 1675.8 | 1649 | - | - | - | 1181 |
| $^{13}$C-$^{13}$C-$^{12}$C | 5 | 1622.5 | 1617 | 5 | 2729.0 | ~1107 | 1142 | 6 | 1672.4 | 1640 | - | - | - | 1157 |
| $^{12}$C-$^{13}$C-$^{13}$C | | | 1617 | | | | 1142 | 7 | 1656.5 | 1637 | - | - | - | 1159 |
| $^{13}$C-$^{13}$C-$^{13}$C | 6 | 1610.5 | 1606 | 6 | 2695.0 | ~1085 | 1120 | 8 | 1651.7 | 1627 | 8 | 2721 | ~1069 | 1136 |

*obtained from corresponding value of ($v_1 + v_3$), without anharmonic correction.

The diagram in Figure 9 (f) compares formation energies as a function of the Fermi level for $C_N=C=C_N$, $C_N$-$C_{Ga}$-$C_N$, and $C_N$. Both tri-C complexes are significantly higher in energy than $C_N$, other defects with one carbon atom, or carbon pair complexes.[10] Although for most Fermi levels the formation energy of $C_N=C=C_N$ complex is ~0.5 eV higher than $C_N$-$C_{Ga}$-$C_N$ configuration, it is nevertheless slightly more stable when the Fermi level is near the valence-band maximum (as shown in Figure 9 (f)). Based solely on formation energies, one would not expect the tri-C complexes would incorporate in GaN. In contrast, the direct comparison of IR absorption intensities of the peaks (MC2; TC1; TC2) demonstrates prevalence of $C_N=C=C_N$ in most of the samples. Hereby, intensities are equal to (28 cm$^{-2}$; 20.2 cm$^{-2}$; 12.7 cm$^{-2}$) at [C]=1.9×10$^{18}$ cm$^{-3}$ and increase up to (119 cm$^{-2}$; 182 cm$^{-2}$; 93 cm$^{-2}$) at [C] = 9×10$^{18}$ cm$^{-3}$. The pronounced incorporation of $C_N=C=C_N$ complexes could be governed by surface kinetic effects unrelated to the equilibrium condition in the bulk crystal considered by DFT. Additionally, while the incomplete decomposition of precursors at the growth interface may be another explanation, samples grown with both butane and pentane result in the same pronounced tri-C modes.

To summarize, models of tri-C defects in the $C_N=C=C_N$ configuration are in excellent agreement with the observed vibrational properties. Figure 9 (f) illustrates that $C_N=C=C_N$ defects exhibit different charge states (from 1+ to 1-) depending on the Fermi level position. It should be noted that the thermodynamic transition levels of $C_N=C=C_N$ defects are closer to mid-gap; the (0/-) level occurs at about 1.5 eV as opposed to 0.9 eV for $C_N$, while (+/0) is located at 1.1 eV and 0.2 eV above VBM, respectively for these defects. Thus it is possible that $C_N=C=C_N$ defects could be responsible for the emission band around 1.62 eV for [C] >10$^{18}$ cm$^{-3}$ reported earlier for the same sample set.[24] At Fermi level positions from 0.9 to 1.1 eV above VBM (incl., Fermi level pinned to $C_N$(0/-) or $C_N=C=C_N$ (+/0)), calculations of the $C_N=C=C_N$ defect predict that the 1+ charge state is stable, indicating that $(C_N=C=C_N)^+$ could compensate negatively charged $C_N$. Since the intensity of vibrational peaks for these tri-C centers is one order of magnitude stronger than for other defects, $C_N=C=C_N$ defects should be considered as an important compensating defect in GaN.

### d. Summary of carbon-related vibrational modes and behavior under illumination

*Table IV. The labels of peaks, detection method (R=Raman, IR = IR absorption), experimental and theoretical vibrational frequencies, and the frequency ratio f for the six point defects discussed in this work.*

| Group | Defect | Label | Method | Exp. ν, cm⁻¹ | f | DFT ν, cm⁻¹ | f | Comment |
|---|---|---|---|---|---|---|---|---|
| Mono-C | $C_N$ | MC1 | R | 766 | 1.028 | 775 | 1.030 | Figure 3 (a) |
| | | MC2 | IR/R | 774 | 1.028 | 785 | 1.030 | Figure 3 (b) |
| | | MC3 | R | 659 | 1.00 | 655 | 1.00 | Figure 3 (c) |
| | | MC4 | R | 673 | 1.00 | 662 | 1.0 | Figure 3 (d) |
| | | MC5 | IR/R | 1540 | 1.028 | - | - | MC2 2nd harmonics |
| | | MC6 | IR/R | 1547 | 1.028 | - | - | |
| | C-H | CH | IR | 2670 | 1.002 | - | - | Stretching mode |
| Di-C | $C_N$-$C_i$ | DC1 | IR/R | 1978 | 1.040 | - | - | Stretching mode, tilt angle ~55° |
| | C≡C? | DC2 | IR | 2250 | 1.040 | - | - | Stretching mode, tilt angle ~30° |
| Tri-C | $C_N=C=C_N$ | TC1 | IR | 1678 | 1.040 | 1671 | 1.040 | Basal config., Figure 9 (e), ($v_3$) |
| | | TC2 | IR/R | 1718 | 1.040 | 1692 | 1.040 | Axial config., ($v_3$) |
| | | TC3 | IR | 2803 | 1.040 | - | 1.040 | Basal config., ($v_1+v_3$) |
| | | TC4 | IR | 2832 | 1.040 | - | 1.040 | Axial config., ($v_1+v_3$) |
| | C=C=C? | TC5 | IR | 1734 | 1.040 | - | - | Another tri-C defect? |

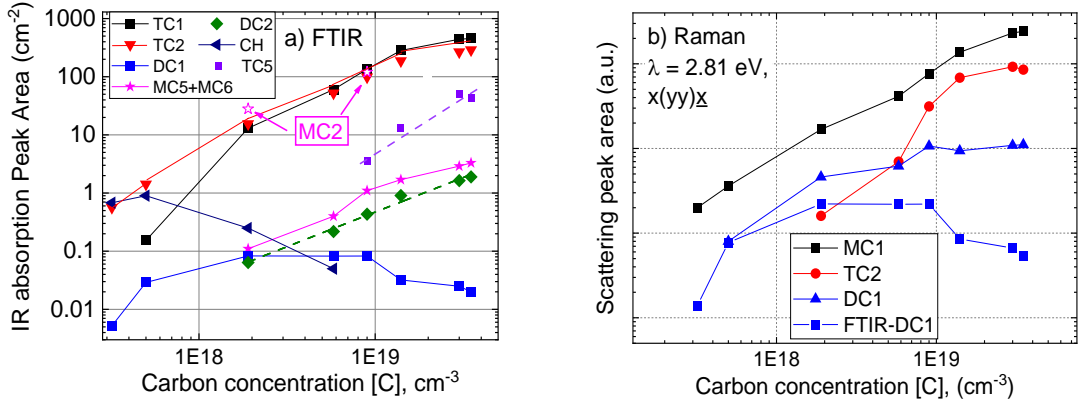

*Figure 10. Integrated peak areas of different vibrational modes at different [C] at room temperature obtained from (a) IR absorption spectra at E⊥c polarization and (b) Raman scattering spectra at 2.81 eV excitation in x(yy)x̲ geometry. FTIR-DC1 signal is reproduced in (b) for comparison with the corresponding Raman mode DC1 (normalized to the value at [C] = 5×10¹⁷ cm⁻³)*

Table IV summarizes 14 vibrational frequencies which we associate with 6 different carbon-containing defects ($C_N$, C-H, $C_N$-$C_i$, C≡C, and $C_N=C=C_N$ (basal and axial)). Table IV includes the vibrational frequencies observed in experiments (IR absorption or Raman (R)) and those ob-tained from DFT calculations. The frequency ratios *f* obtained from the experiments and DFT calculations are shown. For the most pronounced signals of $C_N$ and $C_N=C=C_N$ defects, we refer to the illustrations provided above while for faint vibrational modes we restrict ourselves to the verbal description. Even within the set of samples grown by HVPE in similar conditions, the relation between signals of the defects is distinct at different [C] as shown in Figure 10. Axial $C_N=C=C_N$ defects (TC2) were found in the sample with the lowest doping level, which emphasizes their high prevalence in this material. Although for most samples the direct measurement of IR absorption is difficult in the range of Reststrahlen band, the samples at [C] = $1.9\times10^{18}$ cm$^{-3}$ and [C]=$9\times10^{18}$ cm$^{-3}$ were measured (Figure 1 (b) includes spectrum at [C] = $9\times10^{18}$ cm$^{-3}$) and the absorption of $C_N=C=C_N$ defects (TC1+TC2) is almost twice as intense as the absorption of $C_N$ defects (MC2, 119 cm$^{-2}$). The vibrational modes of $C_N$ defects (MC1-MC4), however, dominate among the other Raman-active defect-related signals. In this respect, we suggest considering these defects as dominating in these samples. We also note that excitation above 2.5 eV (in the range of the defect related UV-Vis absorption)[23] leads to a significant gain in sensitivity in the scattering experiments. Comparing to earlier reports in which excitation at 532 nm (2.33 eV) was used in Raman experiments,[11,27] we get reliable signal detection at 1 – 2 orders of magnitude lower [C]. In our experiments $C_N$ begins to appear at [C] = $3.2\times10^{17}$ cm$^{-3}$.

Note that the relative intensity of the signals is significantly different when using FTIR spectroscopy (Figure 10 (a)) and Raman spectroscopy (Figure 10 (b)) since the incident laser light in the Raman experiments puts the samples into an excited state. The vibrational intensities of photosensitive defects in this case change, and this effect can be observed independently in FTIR experiments with additional light excitation. Figure 11 (a) shows the IR absorption spectra of the most sensitive sample at [C]=$1.9\times10^{18}$ cm$^{-3}$ in the dark and under additional excitation at varied power of a 2.71 eV LED during the measurement of IR absorption. The detector of IR spectrometer is insensitive to this irradiation; we also may exclude the heating of the samples as the temperature-dependent peaks do not change the spectral position and shape.

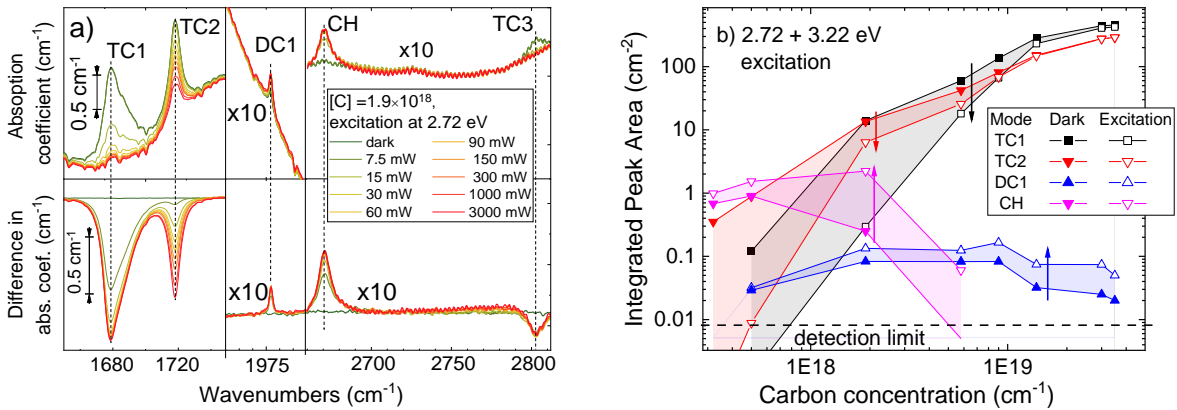

*Figure 11. (a) IR absorption spectra of sample with [C] of $1.9\times10^{18}$ cm$^{-3}$ for E⊥c polarization at room temperature and different powers of additional excitation at 2.72 eV. (b) integrated peak area of different IR absorption peaks at different [C] obtained at combined excitation of 2.72 and 3.22 eV with total power of 4500 mW (optical power of the LEDs, coupling losses are not accounted)*

The optical absorption increases in the sample set at the excitation wavelength from 5 to 20 cm$^{-1}$ with increasing [C], hence, to stimulate the changes of the defects more effectively we include the excitation 3.22 eV (absorption is 5…500 cm$^{-1}$) [22,23]. This excitation, in contrast, has low penetration depth in samples with high [C]. In this respect, the combined excitation (2.71 eV + 3.22 eV) is used for comparison of samples with different [C] (summarized in Figure 11 (b)]). The temperature-sensitive mode TC1 does not change its spectral position or width, so we exclude significant heating of the crystals during the excitation.

Under the excitation at 2.71 eV (or/and 3.22 eV), the signals MC5 and MC6 (not shown), TC1, TC2, TC3, TC4 decrease in intensity, while DC1 and CH become far more pronounced. In addition, CH defects can be present in crystals in the IR inactive state since they are not found in the dark at [C] = 5.8×10$^{18}$ cm$^{-3}$ but appear under excitation (Figure 11(b)). A change in the intensity of vibrational modes can be caused by a change in the defect population (i.e., their charge state changes) or in the structural configuration of these defects. The duration of these processes depends on [C] (and is faster at higher [C]), and ranges from a few seconds to a few minutes. Similar timeframes are typical for the charge transfer observed by Zvanut et. al., or for the bulk photovoltaic effect observed by Levine, Gamov et. al. in C-doped GaN.[19,20,52] One of the mechanisms explaining the simultaneous change of several vibrational modes can be the band-mediated transfer of charge carriers between donor defect and acceptor defect levels, which brings the crystal system out of equilibrium into a metastable excited state.[53]

**Conclusion**

In conclusion, six types of carbon-containing defects were observed by methods of vibrational spectroscopy. The structure of these defects was established via isotope-mass effects analyzed within DFT calculations as well as in the harmonic oscillator approximation for localized vibrational modes. $C_N$ and $C_N=C=C_N$ were identified as pronounced defects with clear signatures in the vibrational spectra and the fraction of $C_N=C=C_N$ increases at high [C]. In Raman experiments, the excitation in the range of defect absorption provided the confident detection of $C_N$ defects even at concentrations as low as 3.2 × 10$^{17}$ cm$^{-3}$. $C_N=C=C_N$ complexes are observed at the same concentration in IR absorption spectra despite their high formation energies from DFT calculations. According to our calculations, these two defects could explain self-compensation during carbon doping in GaN if the Fermi-level position is pinned between the $C_N$ (-/0) and $C_N=C=C_N$ (+/0) transition levels. Other defects, such as CH groups and carbon pairs, are faint in the vibrational spectra of our HVPE crystals but can probably be more prominent for other growth methods, stoichiometry Ga/N, or co-doping. Furthermore, DC1 and CH signals are enhanced by UV-Vis irradiation; therefore, the corresponding defects can exist in the crystal but not appear in the vibrational spectra. Note the importance of these defects - defects showing changes in vibrational modes intensity at additional excitation during FTIR experiments (MC5, MC6, DC1, TC1-TC6, CH) presumably impact the electronic charge balance and thus the conductivity of the material.

# Acknowledgments

The authors would like to thank Prof. Dr. Axel Hoffman and Dr. Harald Scheel from TU Berlin for help with the experiments and valuable discussion.The authors would like to thank Prof. Dr. Axel Hoffman and Dr. Harald Scheel from TU Berlin for help with the experiments and valuable discussion.